\documentclass[american, 12 pt, article]{article}

\usepackage[left=1in, right=1in, top=0.75in, bottom=0.75in]{geometry}
\usepackage[utf8]{inputenc}
\usepackage{setspace}
\usepackage{graphicx}
\usepackage{authblk}
\usepackage{sectsty}
\usepackage{gensymb}
\usepackage{parskip}
\usepackage[labelsep=period,labelfont=bf]{caption}
\usepackage{titlesec}
\usepackage{lmodern} 
\usepackage{wrapfig}
\usepackage[utf8]{inputenc}
\usepackage{chemformula}
\usepackage{siunitx}
\usepackage{xr}
\usepackage[font={small}]{caption}
\usepackage[version=4]{mhchem}
\usepackage{upgreek} 
\graphicspath{{images/}}
\sectionfont{\fontsize{13}{16}\selectfont}

\subsectionfont{\fontsize{12}{14.4}\selectfont}
\subsubsectionfont{\fontsize{11}{13.2}\selectfont}

\titlespacing\section{0pt}{12pt plus 2pt minus 2pt}{12pt plus 2pt minus 2pt}
\usepackage[backend=biber, style=nature] {biblatex}

\usepackage{microtype} 
\usepackage{placeins} 
\usepackage[labelfont=bf]{caption}
\usepackage{amsmath}
\usepackage[version=4]{mhchem} 
\usepackage{upgreek} 
\DeclareSIUnit\angstrom{\text {Å}}

\newcommand{\FexTaS}{\ch{Fe_{\emph{x}}TaS2}}
\newcommand{\HTaS}{2\ch{\emph{H}-TaS2}}
\newcommand{\TTaS}{1\ch{\emph{T}-TaS2}}

\newcommand{\TaS}{\ch{TaS2}}
\newcommand{\HHTaS}{\ch{\emph{H}–TaS2}}

\title{\Large\textbf {Topochemically-engineered coexistence of charge and spin orders in intercalated endotaxial heterostructures}}

\author[1,*]{Samra Husremovi\'{c}}
\author[1]{Wanlin Zhang}
\author[2,3]{Medha Dandu}
\author[4]{Berit H. Goodge}
\author[1]{Isaac M. Craig}
\author[1]{Ellis Kennedy}
\author[1]{Matthew P. Erodici}
\author[5]{Karen C. Bustillo}
\author[5]{Chengyu Song}
\author[5]{Jim Ciston}
\author[2,6]{Sin\'ead Griffin}
\author[2,7]{Archana Raja}
\author[1,7, 8, *]{D. Kwabena Bediako}

\affil[1]{Department of Chemistry, University of California, Berkeley, CA 94720, USA}
\affil[2]{Molecular Foundry, Lawrence Berkeley National Laboratory, Berkeley, CA 94720, USA}
\affil[3]{Present address: Chemistry and Physics of Materials Unit, Jawaharlal Nehru Centre for Advanced Scientific Research, Bengaluru-560064, India}
\affil[4]{\textit{Max-Planck-Institute for Chemical Physics of Solids, N\"{o}thnitzer Str. 40, 01187, Dresden, Germany}}
\affil[5]{National Center for Electron Microscopy, Molecular Foundry, Lawrence Berkeley National Laboratory, Berkeley, CA, USA}
\affil[6]{Materials Sciences Division, Lawrence Berkeley National Laboratory, Berkeley, California 94720, USA}
\affil[7]{\textit{Kavli Energy NanoScience Institute, Berkeley, CA 94720, USA}}
\affil[8]{\textit{Chemical Sciences Division, Lawrence Berkeley National Laboratory, Berkeley, CA 94720, USA}}
\affil[*]{Correspondence to: bediako@berkeley.edu and samra\_husremovic@berkeley.edu}

\date{}
\begin{document}
\maketitle
\doublespacing
\textit{Abstract}\\
\small{Correlated electron systems that host multiple electronic orders offer routes to multifunctional quantum materials, but strong competition between these orders often prevents their coexistence. Here we show that nanoscale, metastable intercalated heterostructures can stabilize a rare combination of long-range magnetism and a commensurate charge density wave (C-CDW) order in a single material. We synthesize a two-dimensional (2D) metastable crystal, \ch{\emph{T/H}-Fe$_x$TaS$_2$}, which comprises an endotaxial polytype heterostructure of \TTaS\ and \HTaS\ with Fe intercalated in the van der Waals interfaces. In \ch{\emph{T/H}-Fe$_x$TaS$_2$}, Fe intercalants provide localized spins that support ferromagnetism, while 1\emph{T} layers host a robust commensurate charge density wave (C-CDW) that persists to room temperature. In these intercalated heterostructures, Fe content simultaneously tunes ordering of spin and charge degrees of freedom, positioning topochemically-prepared intercalated endotaxial heterostructures as a route to stabilize and control competing quantum phases in 2D materials.
}

\newpage
\doublespacing

\section{Introduction}
Investigating correlated systems with multiple electronic orders has been a central focus of academic research. The ultimate goal is to elucidate the interplay and stability of these orders in order to leverage them for technological advancements and new computing paradigms\supercite{nakata2021robust,basov2017towards}. In particular, systems with coexisting quantum phases have garnered great interest due to their potential for multifunctional electronic devices. These systems offer a unique opportunity to encode, store, and retrieve information by exploiting the distinct phases and their interplay\supercite{zhang2022realization,feuer2025charge,kou2018multiferroic,wang2012coexistence,xu2022coexisting}. To this end, transition metal dichalcogenides (TMDs) have been extensively explored due to their ability to host charge density waves (CDWs), superconductivity, and long-range magnetic order in their intercalation compounds\supercite{xie2022structure,wang2020intercalated, hall2019environmental,morosan2006superconductivity,niu2020stabilization,Morosan2007,Hardy2015,Checkelsky2008,Chen2016}. However, these quantum phases often compete: increasing intercalation amount suppresses the nesting conditions that stabilize the CDW phase, while promoting carrier doping and local moments that favor magnetic or superconducting ground states\supercite{maksimovic2022strongly,hall2019environmental,morosan2006superconductivity,niu2020stabilization,wang2020intercalated, chikina2020turning,zhang2022modulation,ni2023crystal,fujisawa2017effect}. 

One approach to stabilizing designer quantum phases is through the formation of metastable materials that are kinetically trapped rather than thermodynamically stable\supercite{Nakajima_2017,horstmann2025dynamic,de2025dynamic,oike2018kinetic,eggebrecht2017light,li2020laser}. Here, we use this strategy to produce two-dimensional crystals that show a rare coexistence of ferromagnetism and a commensurate charge density wave (C-CDW) phase. We synthesize a metastable two-dimensional (2D) crystal, \ch{\emph{T/H}-Fe$_x$TaS$_2$}, comprising a polytype heterostructure of \TTaS\ and \HTaS\ with Fe intercalants occupying the interlayer spacing. These crystals are obtained by treating \TTaS\ flakes with an Fe precursor followed by mild thermal annealing at 250~$^\circ$C. During annealing, Fe is inserted into the van der Waals interface and \TTaS\ layers convert to the thermodynamically favorable \HTaS\ polytype. This yields a mixed-polytype framework and isolates metastable \ch{\emph{T/H}-Fe$_x$TaS$_2$} with coexisting ferromagnetism and C-CDW. This state is not accessible in traditional solid-state synthesis, where growth temperatures above 800~$^\circ$C and intercalation ensure formation of the \HTaS\ polytype and suppression of the C-CDW\supercite{Morosan2007,chen2016charge}.

In \ch{\emph{T/H}-Fe$_x$TaS$_2$}, Fe intercalants provide localized spins that enable ferromagnetism, while the polytype heterostructure hosts the C-CDW phase. Specifically, the \HTaS\ layers act as metallic spacers, while Ta ions within \TTaS\ distort to form 13-atom clusters that arrange periodically in a $\sqrt{13} \times \sqrt{13}$ commensurate superlattice. Recent studies have reported that the C-CDW phase in \emph{T}/\emph{H}\ch{-TaS2} heterostructures is exceptionally robust, persisting even at room temperature\supercite{sung2022two,husremovic2023encoding,sung2024endotaxial}. This intrinsic stability likely enables the C-CDW to withstand the high intercalation levels required to induce ferromagnetism, a coexistence that is highly unconventional in this family of materials. 

Investigating correlated systems with multiple electronic orders has been a central focus of academic research. The ultimate goal is to elucidate the interplay and stability of these orders in order to leverage them for technological advancements and new computing paradigms\supercite{nakata2021robust,basov2017towards}. In particular, systems with coexisting quantum phases have garnered great interest due to their potential for multifunctional electronic devices. These systems offer a unique opportunity to encode, store, and retrieve information by exploiting the distinct phases and their interplay\supercite{zhang2022realization,feuer2025charge,kou2018multiferroic,wang2012coexistence,xu2022coexisting}. To this end, transition metal dichalcogenides (TMDs) have been extensively explored due to their ability to host charge density waves (CDWs), superconductivity, and long-range magnetic order in their intercalation compounds\supercite{xie2022structure,wang2020intercalated, hall2019environmental,morosan2006superconductivity,niu2020stabilization,Morosan2007,Hardy2015,Checkelsky2008,Chen2016}. However, these quantum phases often compete: increasing intercalation amount suppresses the nesting conditions that stabilize the CDW phase, while promoting carrier doping and local moments that favor magnetic or superconducting ground states\supercite{maksimovic2022strongly,hall2019environmental,morosan2006superconductivity,niu2020stabilization,wang2020intercalated, chikina2020turning,zhang2022modulation,ni2023crystal,fujisawa2017effect}. 

One approach to stabilizing designer quantum phases is through the formation of metastable materials that are kinetically trapped rather than thermodynamically stable\supercite{Nakajima_2017,horstmann2025dynamic,de2025dynamic,oike2018kinetic,eggebrecht2017light,li2020laser}. Here, we use this strategy to produce two-dimensional crystals that show a rare coexistence of ferromagnetism and a commensurate charge density wave (C-CDW) phase. We synthesize a metastable two-dimensional (2D) crystal, \ch{\emph{T/H}-Fe$_x$TaS$_2$}, comprising a polytype heterostructure of \TTaS\ and \HTaS\ with Fe intercalants occupying the interlayer spacing. These crystals are obtained by treating \TTaS\ flakes with an Fe precursor followed by mild thermal annealing at 250~$^\circ$C. During annealing, Fe is inserted into the van der Waals interface and \TTaS\ layers convert to the thermodynamically favorable \HTaS\ polytype. This yields a mixed-polytype framework and isolates metastable \ch{\emph{T/H}-Fe$_x$TaS$_2$} with coexisting ferromagnetism and C-CDW. This state is not accessible in traditional solid-state synthesis, where growth temperatures above 800~$^\circ$C and intercalation ensure formation of the \HTaS\ polytype and suppression of the C-CDW\supercite{Morosan2007,chen2016charge}.

In \ch{\emph{T/H}-Fe$_x$TaS$_2$}, Fe intercalants provide localized spins that enable ferromagnetism, while the polytype heterostructure hosts the C-CDW phase. Specifically, the \HTaS\ layers act as metallic spacers, while Ta ions within \TTaS\ distort to form 13-atom clusters that arrange periodically in a $\sqrt{13} \times \sqrt{13}$ commensurate superlattice. Recent studies have reported that the C-CDW phase in \emph{T}/\emph{H}\ch{-TaS2} heterostructures is exceptionally robust, persisting even at room temperature\supercite{sung2022two,husremovic2023encoding,sung2024endotaxial}. This intrinsic stability likely enables the C-CDW to withstand the high intercalation levels required to induce ferromagnetism, a coexistence that is highly unconventional in this family of materials. 


\section{Synthesis of Fe-intercalated polytype heterostructures}

\begin{figure*}[!htbp]
    \centering
    \includegraphics[width=\textwidth]{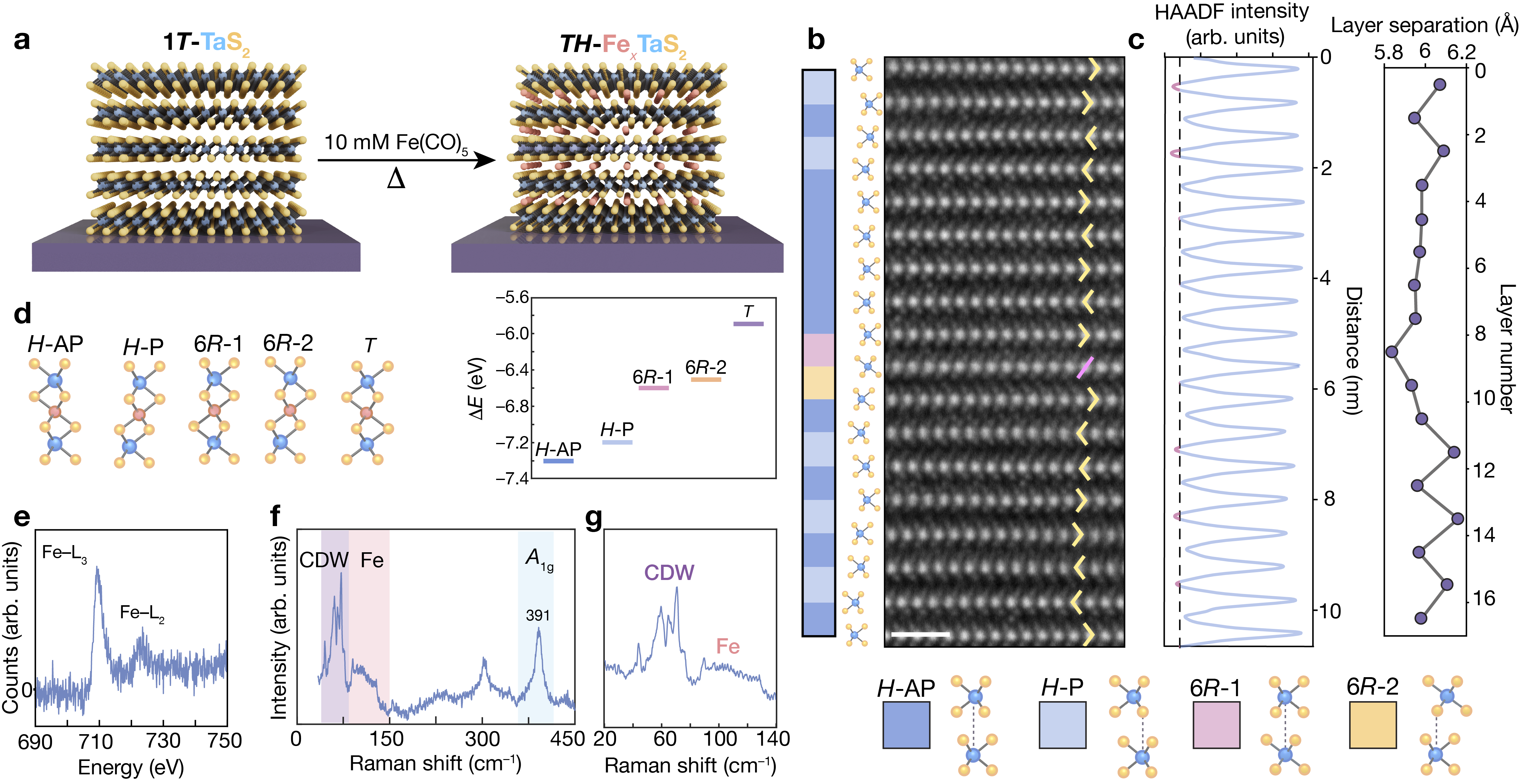}
    \caption[Synthesis of Fe-intercalated endotaxial heterostructures.] {Synthesis of Fe-intercalated endotaxial heterostructures. \textbf{(a)} Schematic depiction of synthesizing Fe-intercalated polytype heterostructures (\ch{\emph{T/H}-Fe$_x$TaS$_2$}) from two-dimensional (2D) \TTaS\ crystals. \textbf{(b)} Room-temperature atomic-resolution HAADF-STEM micrograph of a cross-sectioned \ch{\emph{T/H}-Fe$_x$TaS$_2$} heterostructure along the [$10\overline{1}0$] zone axis. Yellow and pink markings on the HAADF image signify \HHTaS\ and \TTaS\ layers, respectively. The atomic model of the imaged layers is depicted on the left, with adjacent colors signifying the type of crystallographic stacking according to the legend below the HAADF image. The Fe/Ta ratio of the crystal is 0.17(2) from STEM-EDS analysis. Scale bar: 1 nm. \textbf{(c)} Left: Averaged vertical line profile of the image in (b) obtained along the direction of the layers. A dashed line is placed at the HAADF intensity measured for a fragment of \textit{H}--AP layers from the top of the heterostructure. Right: Interlayer distance across the heterostructure in (b). \textbf{(d)} Structures of distinct interlayer stacking registries intercalated with Fe and their corresponding change in energy upon intercalation ($\Delta E$) calculated using DFT. Here, $\Delta E$ corresponds to the total change in energy between a $2 \times 2$ supercell slab geometry of Fe${1}$Ta${8}$S$_{16}$ and the corresponding bare lattice, where only the Fe position is relaxed. \textbf{(e)} Cumulative EEL spectrum of the sample from (b) acquired at liquid nitrogen temperature. \textbf{(f)} Raman spectrum of \ch{\emph{T/H}-Fe$_x$TaS$_2$} crystal from (b) prior to cross-sectioning. Data is collected on a solid \ch{Si3N4/Si} support at room temperature. The spectral regions where dominant modes originate from the Fe superlattice, CDW supercell, and the host-lattice $A_{1g}$ mode \supercite{fan2021excitations,husremovic2023encoding,husremovic2022hard} are highlighted. \textbf{(g)} Raman spectrum (f) in the low-frequency range. The sharp CDW-related modes and the broader Fe-related modes are marked.
    }
    \label{Fig1}
\end{figure*}

To synthesize \ch{\emph{T/H}-Fe$_x$TaS$_2$}, we mechanically exfoliate \TTaS\ crystals into thin flakes, treat them with a carbonyl-derived Fe precursor, and then subject them to mild thermal annealing under vacuum at 250~$^{\circ}$C (Figure~\ref{Fig1}a). This mild annealing step induces two concurrent processes: (1) a partial polytype transition from \TTaS\ to \HHTaS\ and (2) intercalation of Fe from the precursor into the interlayer spacing of TaS$_2$\supercite{sung2022two,husremovic2023encoding,husremovic2025tailored}. The result is an intercalated polytype heterostructure, \ch{\emph{T/H}-Fe$_x$TaS$_2$}, comprising $T$ and $H$ polytype layers with Fe ions residing in the van der Waals gaps. Notably, this metastable heterostructure has no bulk analogue, as bulk intercalated TMDs are synthesized at high temperatures that stabilize the thermodynamically favored Fe-intercalated 2$H$-TaS$_2$\supercite{Morosan2007}.

To examine the structure of \ch{\emph{T/H}-Fe$_x$TaS$_2$}, we cross-sectioned 2D crystals and imaged them with atomic-resolution high-angle annular dark-field scanning transmission electron microscopy (HAADF-STEM), as seen in Figure \ref{Fig1}b. HAADF-STEM micrographs revealed different interlayer stacking registries with homotypic and heterotypic polytype interfaces (Figure \ref{Fig1}b). The observed heterogeneity in interlayer stacking can be attributed to interlayer slips during the polytype transformation process, which serve to prevent the direct overlap of sulfur ions in adjacent layers\supercite{husremovic2023encoding}. Notably, all detected interlayer registries contain pseudo-octahedral interstitial sites amenable to intercalation (Figure~\ref{Fig1}b). These sites have slightly different local geometries across the stacking registries because the interlayer spacing varies with stacking (Figure~\ref{Fig1}c, right). Despite these geometric differences, pseudo-octahedral sites at all interfaces host intercalants, as evidenced by the non-zero HAADF–STEM intensity between the layers at every interface (Figure~\ref{Fig1}b,c).

Interestingly, we observe that the HAADF intensity is generally lower between the parallel-stacked \HHTaS\ layers (\textit{H}--P) compared to the antiparallel-stacked \HHTaS\ layers (\textit{H}--AP), which could suggest a preferential intercalation in the latter interfaces (Figure \ref{Fig1}b,c). Antiparallel \HHTaS\ stacking (\textit{H}--AP) is found in bulk intercalated \textit{H}-\FexTaS\ crystals, indicating that it is the thermodynamically favorable configuration\supercite{Morosan2007,Checkelsky2008,Hardy2015,husremovic2022hard}. In the \textit{H}--AP configuration, the adjacent layers of \HHTaS\ are rotated by 180$^{\circ}$, leading to a direct alignment of the tantalum (Ta) ions across the trigonal prismatic interfaces. On the other hand, in the parallel \HHTaS\ stacking (\textit{H}--P) configuration, adjacent layers are identical but translationally offset\supercite{VanWinkle2023}. This yields a configuration where the metals are aligned with sulfur ions across the trigonal prismatic interfaces\supercite{VanWinkle2023}. We calculated the intercalation energy of all experimentally observed stacking configurations and the homotypic \TTaS\ interfaces using density functional theory (DFT).  These calculations confirmed that intercalated \textit{H}--AP interfaces are the most thermodynamically favorable, with the highest energy stabilization upon intercalation (Figure \ref{Fig1}d). Moreover, DFT reveals that intercalation is generally less favorable for interfaces comprising \TTaS\ layers (Figure \ref{Fig1}d). These results indicate that intercalation destabilizes the \TTaS\ phase, explaining why bulk Fe-intercalated \TTaS\ has been  elusive\supercite{niu2020stabilization}. In addition, the destabilization of \TTaS\ upon intercalation is congruent with our experimental observations in these topotactic reactions; thermal annealing of \TTaS\ following chemical treatment generally yields a considerably larger change in optical contrast—a highly reproducible indicator of polytype transitions in \TaS\supercite{husremovic2023encoding}—compared to annealing in the absence of the Fe intercalant precursor (Supplementary Figure 1). However, unlike conventional high-temperature crystal growth, we find that metastable Fe-intercalated configurations comprising both \HHTaS\ and \TTaS\ can be realized by annealing the chemically treated crystals for short durations and at mild temperatures around 250$^{\circ}$C.

We supplement our atomic-resolution imaging with elemental composition analysis using electron energy loss spectroscopy (EELS) and STEM paired with energy dispersive X-ray spectroscopy (STEM-EDS). This elemental analysis is crucial as the intercalants could be `self-intercalated' Ta\supercite{Bonilla2020} and not the desired Fe intercalants. EEL spectra show distinct features of Fe (Figure \ref{Fig1}e), while STEM-EDS reveals a Fe-to-Ta ratio of 0.17(2) in this crystal. The presence of Fe is also evident in the Raman spectra of \textit{T/H}-\FexTaS, which display a strong $A_{\mathrm{1g}}$ mode near 390~cm$^{-1}$, characteristic of \FexTaS\ with disordered intercalants\supercite{husremovic2022hard} (Figure~\ref{Fig1}f,g). In pristine \HTaS, the $A_{\mathrm{1g}}$ mode appears around $\sim$400~cm$^{-1}$ and shifts toward $\sim$390~cm$^{-1}$ upon electron doping during Fe intercalation\supercite{husremovic2022hard,fan2021excitations}, whereas \TTaS\ exhibits its $A_{\mathrm{1g}}$ mode near 380~cm$^{-1}$\supercite{husremovic2023encoding,lacinska2022raman}. The $A_{\mathrm{1g}}$ frequency observed in our intercalated polytype heterostructure is therefore consistent with a dominant Fe-intercalated H-polytype.

Intriguingly, we also observe sharp Raman modes in the ultralow frequency (ULF) region, characteristic of the commensurate CDW phase (C-CDW) in \TTaS\ layers (Figure \ref{Fig1}g). Although in unintercalated polytype heterostructures, the C-CDW phase exhibits enhancement compared to \TTaS—with its ordering temperature increasing from below 180 K to above room temperature\supercite{sung2022two,sung2024endotaxial,sung2024endotaxial}—the persistence of C-CDW enhancement in intercalated samples is surprising. Intercalation typically eliminates C-CDW ordering due to structural and charge perturbations engendered by intercalants\supercite{niu2020stabilization,hall2019environmental,morosan2006superconductivity,niu2020stabilization,wang2020intercalated}. Because our atomic-resolution imaging confirms that Fe is present in the interfaces that comprise \TTaS\ (Figure \ref{Fig1}b,c), this raises important questions about the nature of co-localization versus segregation of the domains of Fe intercalants and CDW phases within the $ab$ plane.

\section{Coexistence of Fe and CDW superstructures}
\begin{figure*}[!htbp]
    \centering
    \includegraphics[width=\textwidth]{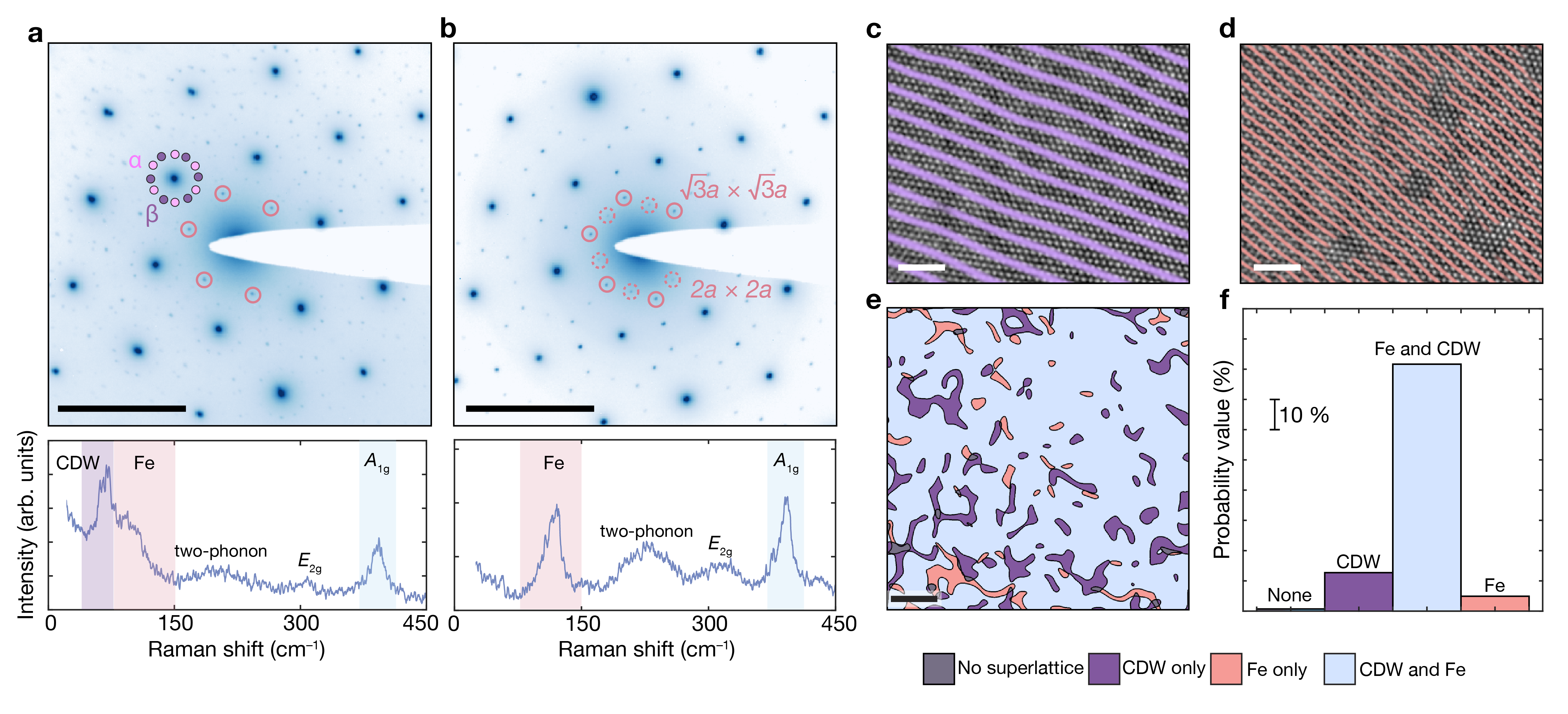}
    \caption[Coexistence of intercalant and CDW superlattices in \ch{\emph{T/H}-Fe$_x$TaS$_2$} heterostructures.] {Coexistence of intercalant and CDW superlattices in \ch{\emph{T/H}-Fe$_x$TaS$_2$} heterostructures. \textbf{(a,b)} Raman spectra and corresponding SAED patterns of two \ch{\emph{T/H}-Fe$_x$TaS$_2$} heterostructures on electron-transparent \ch{Si3N4} TEM grids. For each sample, Raman spectra and SAED patterns were collected in nearly identical regions of the crystal. In SAED patterns, the first-order Fe superlattice and CDW superstructure reflections are marked. The strongly reflecting spots originate from the host lattice (\textit{TH}-\ch{TaS2}). In the Raman spectra, spectral regions where phonon modes for different superlattices are expected\supercite{fan2021excitations,husremovic2023encoding,husremovic2022hard} are highlighted. The crystal in (a) exhibits both CDW- and Fe-related peaks, whereas the sample in (b) exhibits only Fe-related peaks. \textbf{(c,d)} Atomic-resolution plan-view ($\parallel$ \textit{c}-axis) HAADF-STEM image of the heterostructure from (a) overlaid with the inverse FFT (iFFT) image generated from a CDW superlattice peak \textbf{(c)} and a Fe superlattice peak \textbf{(d)}. Scale bars (c,d): 2 nm. \textbf{(e)} Superlattice domain map of sample from (a) generated using FFT analysis. Scale bar: 5 nm. \textbf{(f)} Histogram of the domain map from (e).      
    }
    \label{Fig2}
\end{figure*}

To investigate whether Fe and CDW structural order coexist and whether CDW domains are absent where Fe intercalants are present, we examined in-plane inhomogeneity by preparing samples for plan-view imaging (parallel to the $c$ axis). We first carried out selected area electron diffraction (SAED) and Raman spectroscopy on nearly identical regions (within 1 $\mu$m) of \TTaS\ crystals that underwent chemical treatment and annealing at 250 $^{\circ}$C for 1.5 hours. These measurements revealed two distinct classes of crystals, with representative structural analyses shown in Figure~\ref{Fig2}a,b. 

In the first class, SAED patterns display C-CDW chiral mirror domains ($\alpha$ and $\beta$) consistent with the CDW structure of unintercalated polytype heterostructures (Figure~\ref{Fig2}a)\supercite{husremovic2023encoding,sung2024endotaxial}. Superimposed on these CDW superstructures, we also observe faint $\sqrt{3} \times \sqrt{3}$ Fe superlattice reflections, indicative of clustered, partially disordered intercalants (Figure~\ref{Fig2}a). Notably, $\sqrt{3} \times \sqrt{3}$ peaks are absent in pristine \ch{TaS2} and appear only upon intercalation, which makes them a clear indicator of ordered Fe species\supercite{husremovic2022hard,husremovic2025tailored,fan2021excitations}. While this superlattice periodicity is commonly associated with a high Fe content near Fe/Ta $\approx$ 0.33, recent studies show that the same ordering can also emerge at lower intercalation levels, as intercalants cluster into $\sqrt{3} \times \sqrt{3}$ domains\supercite{craig2025modeling,husremovic2025tailored}. Raman spectra were obtained in the same regions as SAED (Figure~\ref{Fig2}a,b). These spectra exhibit sharp C-CDW modes in the ultralow-frequency (ULF) range together with broad Fe-related modes (Figure~\ref{Fig2}a, bottom). 

The second class of crystals shows only Fe-related superstructures in both SAED and Raman data (Figure~\ref{Fig2}b). In these crystals, Fe-related Raman modes are sharper and blue-shifted, consistent with higher Fe concentrations than in the first class (Figure~\ref{Fig2}a,b). Together, these observations indicate that higher Fe intercalation drives a transformation to \textit{H}--\FexTaS\ and suppresses the C-CDW phase originally hosted by the \TTaS\ layers. In contrast, lower Fe concentrations permit coexistence of Fe and C-CDW ordering within the spatial resolution of our SAED and Raman probes (700 nm -- 1 $\mu$m).

SAED and Raman measurements probe CDW and Fe ordering but cannot exclude phase segregation at the atomic scale. To address this limitation, we acquired atomic-resolution HAADF–STEM images along the $c$-axis from crystals that display both CDW and Fe ordering in SAED. Fourier transforms of these images show that C-CDW and Fe superlattice fringes frequently coexist within the same regions (Figure \ref{Fig2}c,d). Breaks in periodicity, indicative of local disorder, are present in both modulations but do not exhibit discernible correlation (Figure \ref{Fig2}c--f). The most common structural motif is instead concurrent CDW and Fe ordering (Figure \ref{Fig2}e,f). Although plan-view imaging does not capture out-of-plane disorder, taken together with our cross-sectional data (Figure \ref{Fig1}b), these observations demonstrate that CDW and Fe order can occupy the same real-space regions and establish a structural basis for magnetoelectronic coupling between them.

\section{Effects of intercalation on long-range spin and charge ordering}
\begin{figure*}[!htbp]
    \centering
    \includegraphics[width=\textwidth]{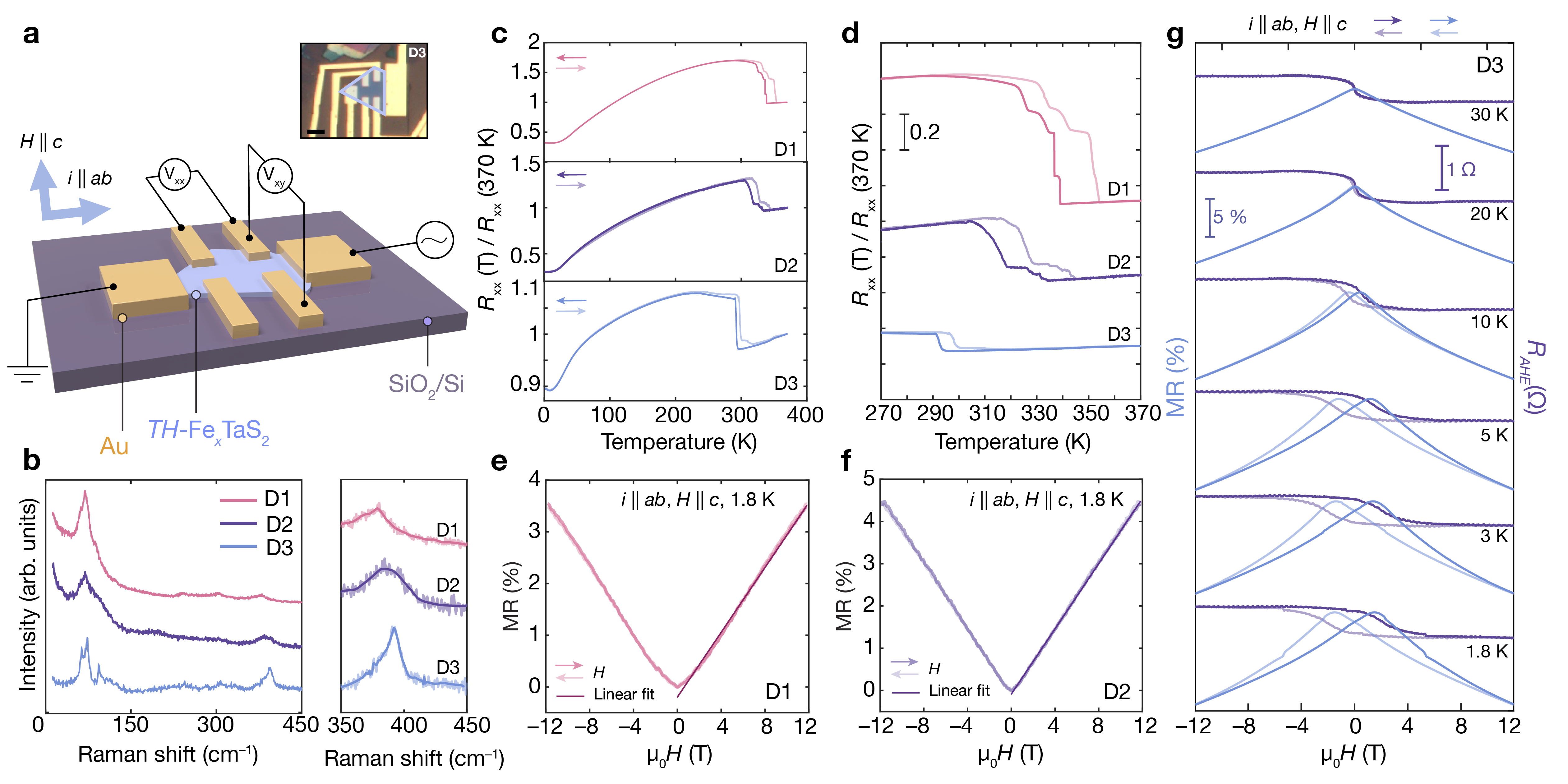}
    \caption[Electronic evidence of spin and charge ordering in \textit{TH}-\ch{TaS2} heterostructures.] {Electronic evidence of spin and charge ordering in \textit{TH}-\ch{TaS2} heterostructures. \textbf{(a)} Schematic of a \ch{\emph{T/H}-Fe$_x$TaS$_2$} device on \ch{SiO2/Si}. Inset: Optical micrograph of device D3, with the sample outline marked in light blue and the sample area false-colored. Scale bar: 2 $\mu$m. \textbf{(b)} Raman spectra of measured devices, labeled D1---D3. Device D1 is a pristine polytype heterostructure (\textit{TH}-\ch{TaS2}) without Fe. Device D2 is lightly intercalated, with an Fe/Ta ratio ($x$) of 0.02(1) determined from STEM-EDS. Device D3 exhibits Raman features consistent with $\sim 0.15 < x < 0.2$. \textbf{(c)} Temperature-dependent normalized resistance ($R_{xx}(T) / R_{xx}(370 K)$) for devices D1--D3. \textbf{(d)} High-temperature segment of (c). Curves are vertically shifted for clarity. \textbf{(e,f)} Magnetoresistance of D1 \textbf{(e)} and D2 \textbf{(f)} at 1.8 K. Linear fit to the data is shown. \textbf{(g)} Magnetoresistance (MR) and field-dependent anomalous Hall resistance ($R_{\mathrm{AHE}}$) of device D3 as a function of temperature. Curves are vertically offset for clarity.
    }
    \label{Fig3}
\end{figure*}

To elucidate how intercalation modifies magnetism and charge ordering in \textit{T/H}--\FexTaS\ crystals, we conducted variable-temperature transport measurements on Fe-intercalated polytype heterostructures (Figure \ref{Fig3}a). Three representative devices, D1–D3, are examined. D1 serves as a reference polytype heterostructure without Fe, whereas D2 and D3 are intercalated, as evidenced by their broad Fe-related Raman modes and a blue shift of the $A_{\mathrm{1g}}$ mode relative to D1 (Figure \ref{Fig3}b). These Raman shifts are more pronounced in D3 (Figure \ref{Fig3}b)\supercite{husremovic2022hard}, consistent with a higher Fe concentration than in D2. All devices exhibit Raman active CDW-related phonon modes (Figure \ref{Fig3}b) and sharp resistance changes above 280 K that mark CDW transitions from the low-resistance incommensurate (IC) phase to the high-resistance C-CDW phase\supercite{scruby1975role,sung2022two,husremovic2023encoding} (Figure \ref{Fig3}c,d). The CDW transition temperatures decrease systematically from D1 to D3 (Figure \ref{Fig3}c,d), in parallel with increasing Fe content. These data indicate that Fe intercalation does frustrate the CDW order, but does not eliminate this phase entirely.

Having established the persistence of charge ordering in D1--D3, we now turn to examining whether these devices exhibit long-range spin ordering (magnetism). To this end, we measured their magnetoresistance profiles at cryogenic temperatures (Figure \ref{Fig3}e--g). Magnetoresistance curves of D1 and D2 exhibit similar properties: nearly linear positive magnetoresistance and overlapping forward and backward field sweeps (Figure \ref{Fig3}e,f), consistent with a lack of long-range magnetic order. Conversely, D3 exhibits hysteretic magnetoresistance and an anomalous Hall effect up to 30 K, consistent with ferromagnetism (Figure \ref{Fig3}g)\supercite{husremovic2022hard}. The hysteretic profiles in this device are broad, pointing to intercalant disorder. However, the coercivities remain high, exceeding 1.5 T at 1.8 K (Figure \ref{Fig3}g). We note that at cryogenic temperatures, the transport properties become dominated by the metallic \HTaS\ layers. For this reason, CDW ordering cannot be electronically assessed concurrently with ferromagnetism, mandating another probe to establish whether CDW ordering also persists at low temperatures.


\section{Coexistence of CDWs and magnetism}
\begin{figure*}[!htbp]
    \centering
    \includegraphics[width=4.5in]{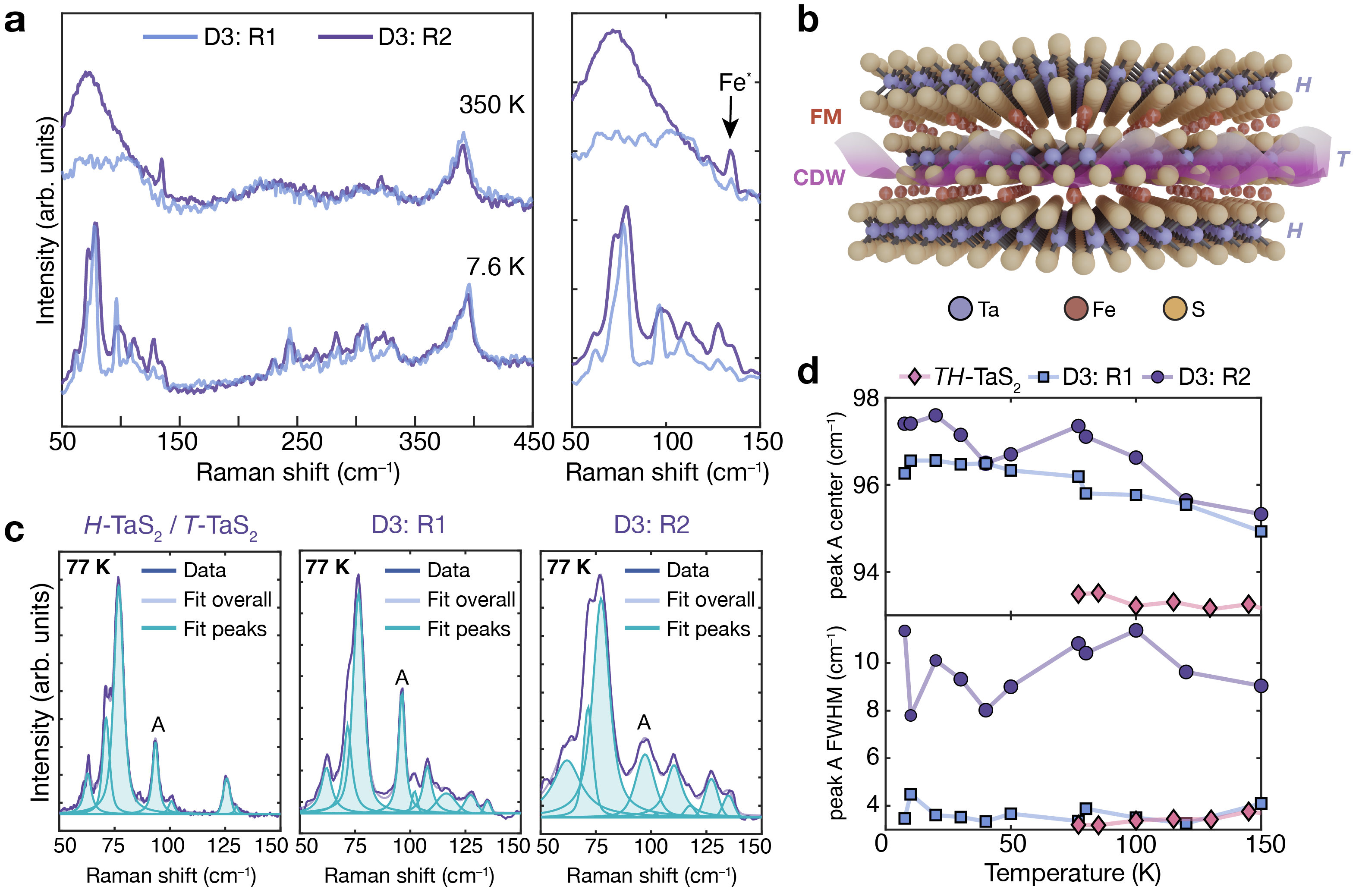}
    \caption[Charge density wave ordering of the ferromagnetic device D3.] {Charge density wave ordering of the ferromagnetic device D3. \textbf{(a)} Raman spectra of device D3 acquired at 7.6 K and 350 K in two regions: (1) the main channel area (region R1) and (2) near the device edge (region R2). The right panel shows the low-frequency portion of the spectra, with the iron-related (Fe\textsuperscript{*}) mode at $\sim$134.5~cm$^{-1}$ indicated by an arrow. \textbf{(b)} Schematic illustration of an intercalated polytype heterostructure, where ferromagnetism (FM) arises from Fe intercalants, and the \TTaS\ layers host a CDW phase. \textbf{(c)} Low-frequency Raman spectra of (1) a pristine polytype heterostructure, (2) device D3 region 1, and (3) device D3 region 2 at 77 K. Pseudo-Voigt fits to the data are shown. Spectra of the pristine heterostructure and device D3 were acquired using 633 nm and 532 nm lasers, respectively. Optical contrast measurements\supercite{husremovic2023encoding} reveal that the pristine polytype heterostructure comprised five \textit{H}-\ch{TaS2} layers out of a total of 20 \ch{TaS2} layers. \textbf{(d)} Temperature-dependent peak position and full width at half maximum (FWHM) of Raman mode A, labeled in (c). Peak parameters were obtained from pseudo-Voigt fits. 
    }
    \label{Fig4}
\end{figure*}

To examine whether the CDW order coexists with ferromagnetism, we probed the CDW structure of D3 at temperatures below its magnetic ordering transition. Two regions of D3 were investigated. One was the channel probed in the transport measurements of Figure \ref{Fig3}g (region R1) and the other was an area near the sample edge (region R2). Representative Raman spectra at 7.6 K and 350 K are shown in Figure \ref{Fig4}a. At 7.6 K, in the ferromagnetically ordered phase, both R1 and R2 exhibit sharp Raman modes below 150 cm$^{-1}$ that are consistent with a C-CDW phase. Together with the transport data, these measurements demonstrate that ferromagnetism and C-CDW order coexist in D3 (Figure \ref{Fig4}b). At 350 K, the CDW modes broaden and become more disordered, congruent with melting of the C-CDW phase. In this high-temperature regime, the CDW modes are sufficiently broad that the Fe-related mode at $\sim$134.5 cm$^{-1}$ is spectrally isolated from neighboring CDW modes, allowing a direct comparison of Fe ordering in R1 and R2. Both regions evince a Fe-related peak, but its intensity is markedly higher in R2 (Figure \ref{Fig4}a, right panel), indicating a higher concentration of ordered Fe in this region, likely reflecting a higher local Fe content.

Next, we investigate whether differences in Fe ordering between R1 and R2 correlate with differences in their CDW-related Raman modes. To this end, we compared Raman spectra collected at 77 K from (1) a reference polytype heterostructure without Fe and (2) regions R1 and R2 of D3 (Figure \ref{Fig4}c, Supplementary Figure 2). Our analyses focus on a representative, well-isolated C-CDW mode (peak A in Figure \ref{Fig4}c). In the intercalated regions, peak A is systematically blue-shifted relative to the pristine heterostructure (Figure \ref{Fig4}d, top panel), consistent with bond stiffening due to the presence of intercalants\supercite{fan2021excitations,tsang1978raman}. Pseudo-Voigt analysis of peak A shows that its full width at half-maximum (FWHM) is similar in the pristine heterostructure and in R1 (Figure \ref{Fig4}d, bottom panel). In contrast, the FWHM is markedly higher for R2, which has the highest concentration of ordered Fe. Taken together, these observations indicate an inverse correlation between Fe and CDW ordering: regions with stronger Fe order exhibit more disordered CDW signatures. We note that intercalation-induced CDW defects may be complex. These may induce CDW phase slips (introduction of discommensurations) or CDW amplitude changes\supercite{mutka1983pinning}.  Moreover, these changes may engender asymmetry of CDW clusters\supercite{fujisawa2017effect,zhang2022visualizing}, irregularities in the C-CDW superlattice\supercite{lutsyk2023influence, fujisawa2017effect,zhang2022visualizing}, metallic defects, and narrow metallic domain walls\supercite{zhang2022visualizing}. 

\begin{figure*}[!htbp]
    \centering
    \includegraphics[width=4.5in]{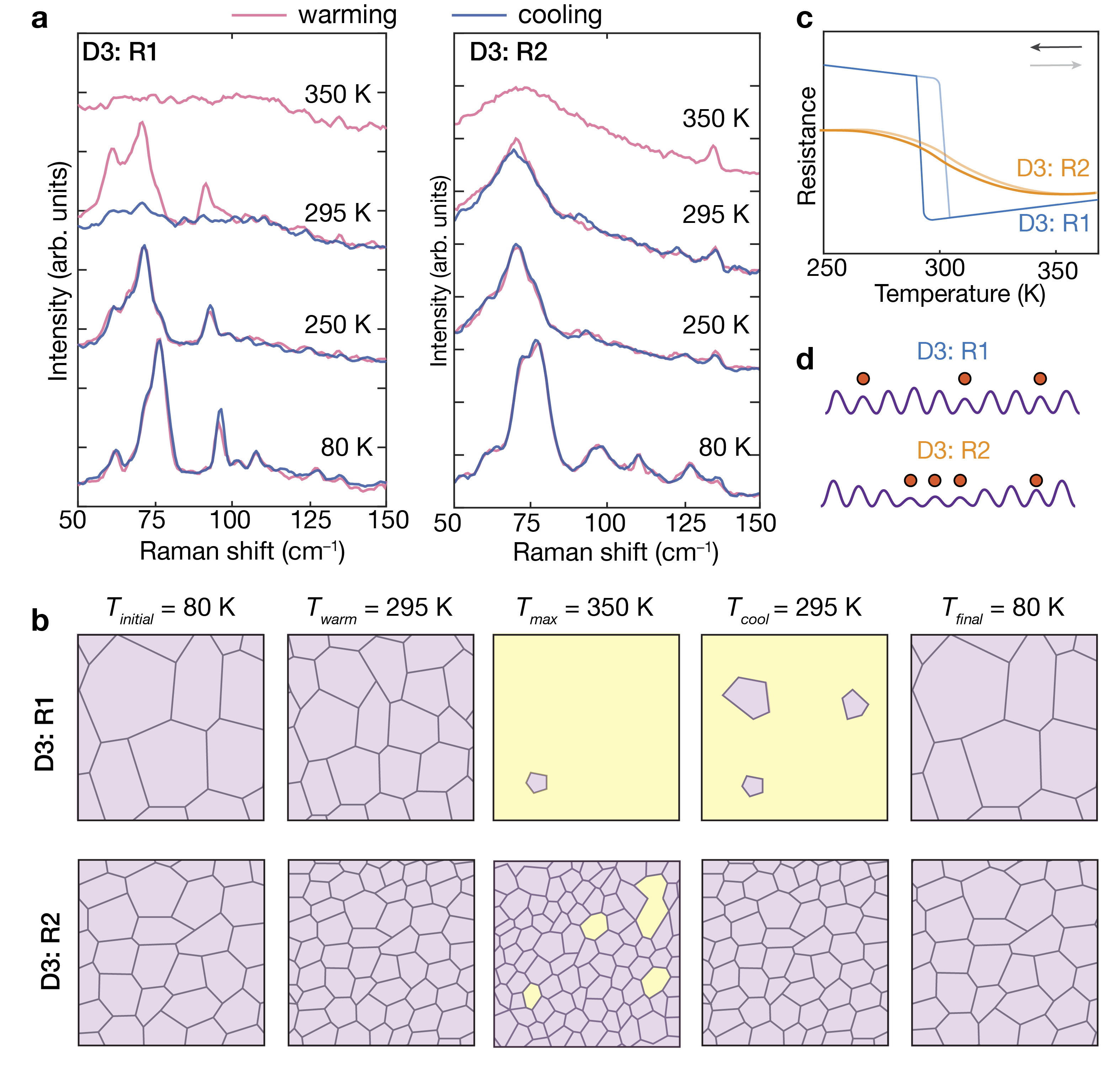}
    \caption[Effects of intercalant ordering on the CDW thermal hysteresis.] {Effects of intercalant ordering on the CDW thermal hysteresis. \textbf{(a)} Raman spectra measured upon cooling from 350 K (blue) and warming from 7.6 K (red) for device D3 in regions R1 (main measurement channel) and R2 (edge of the device). Data was collected using a 532 nm laser. \textbf{(b)} Cartoon illustration of the potential temperature-dependent domain evolution for D3: R1 and D3: R2. Purple lines represent disordered CDW domains and domain walls. Violet and yellow regions represent the C-CDW and IC-CDW domains, respectively. \textbf{(c)} Hypothesized resistivity curves for regions R1 and R2, inferred from the Raman data. Curve for R1 matches the measured data in Figure 3. \textbf{(d)} Illustration showing that Fe intercalants (orange circles) disrupt the CDW order (violet curve), with regions R1 and R2 exhibiting different Fe distributions that yield distinct patterns of CDW disorder.}
    \label{Fig5}
\end{figure*}

Subsequently, we examine how the temperature-dependent CDW response differs between regions of D3. Figure \ref{Fig5}a shows Raman spectra from R1 and R2 collected (1) upon warming from 80 K to 350 K and (2) upon cooling from 350 K to 80 K. Region R1 exhibits pronounced thermal hysteresis near 295 K: CDW peaks are well defined upon warming and largely absent upon cooling. This behavior mirrors the electronic transport data in Figure \ref{Fig3}d. In contrast, CDW modes in R2 do not show noticeable hysteresis near room temperature, and their intensities at 350 K are significantly higher than those of R1. This suggests that temperatures above 350 K would be required to fully melt the C-CDW domains in R2. Such robustness against thermal cycling is consistent with strong pinning by ordered Fe intercalants\supercite{mutka1983pinning,fazekas1979electrical}, which act as defects that hinder the evolution of CDW domains (Figure \ref{Fig5}b).
A schematic of the proposed CDW domain evolution in R1 and R2 is shown in Figure \ref{Fig5}b. We propose that in R2, defects pin CDW domains and prevent them from fully melting into the IC-CDW phase up to at least 350 K. In this scenario, the corresponding thermal hysteresis loop is broader and shifted relative to that of R1, as illustrated in Figure \ref{Fig5}c. Thus, our data indicate that the distinct Fe configurations in R1 and R2 underpin their contrasting CDW behavior (Figure \ref{Fig5}d). In R2, enhanced Fe ordering coincides with increased CDW disorder and an expanded thermal hysteresis window.

\section{Conclusions and Outlook}
In this work, we demonstrate the synthesis of \ch{\emph{T/H}-Fe$_x$TaS$_2$}, a metastable Fe-intercalated polytype heterostructure. This material is prepared by treating \TTaS\ flakes with solutions of iron pentacarbonyl, followed by mild thermal annealing at 250~$^{\circ}$C. This approach enables the realization of coexisting \HHTaS\ and \TTaS\ layers intercalated with Fe, an 2D intercalation compound with no bulk analogue. In this structure, Fe ions provide localized spins for magnetism, while the \TTaS\ layers of the polytype heterostructure host a robust CDW. We find that domains of structurally ordered Fe can co-localize with C-CDW domains down to the nanometer scale. At sufficiently high Fe concentration, we also observe an onset of ferromagnetism that is maintained despite a coexisting C-CDW phase. Increasing Fe concentration further introduces disorder into the CDW phase and decreases the CDW transition temperature. Nevertheless, the C-CDW phase remains robust up to room temperature, even in the presence of sufficiently high amounts of Fe that allow for magnetism. Interestingly, we discover that Fe ordering can pin CDW domains in crystals with nominally higher CDW disorder, creating a metastable “trapped” CDW state. This state allows the crystals to maintain their low-temperature CDW phase even at temperatures exceeding 350 K, where CDW domains in more ordered crystals typically melt. This work shows how topotactic modification of 2D materials—such as the introduction of magnetism by intercalation of a CDW material—can be used to stablize unconventional hybrid correlated electron systems that provide unparalleled opportunities to study and exploit the interplay between exotic, typically competing, quantum phases.

\section{Methods}
\subsection{Synthesis of bulk \TTaS\ crystals}
Single crystals of \TTaS\ were grown by chemical vapor transport (CVT) using powders of elemental Ta (–100 mesh, 99.98$\%$, Nb 50 ppm, Alfa Aesar), S (99.999$\%$, Acros Organics), and \ch{I2} (99.999$\%$, Spectrum Chemicals), all used as received. Sulfur and tantalum precursors were loaded in a 2.1:1 stoichiometry together with 5 mg/cm$^3$ of \ch{I2} into a fused quartz ampule (14 mm inner diameter, 1 mm wall thickness, 12 cm length) under vacuum ($\sim$$1 \times 10^{-5}$ Torr). The ampule was placed in a two-zone tube furnace with the hot zone maintained at 1050~$^\circ$C and the cold (growth) zone at 950~$^\circ$C for two weeks, and then quenched from this temperature profile into ice water.

\subsection{\TTaS\ exfoliation}
Substrates (90 nm SiO$_2$/Si, Nova Electronic Materials) were first cleaned using oxygen plasma for 2-3 minutes and then baked on a hotplate in an Ar glovebox for approximately 1 hour to remove adsorbates. Immediately before exfoliation, a $\sim 2 \times 2$ mm fragment of a \TTaS\ crystal was repeatedly cleaved using Magic Scotch tape. The cleaned substrates were then placed onto the tape, held under gentle finger pressure for 10 minutes, and rapidly peeled away to transfer thin \TTaS\ flakes onto the SiO$_2$/Si surface.

\subsection{Synthesis of \ch{\emph{T/H}-Fe$_x$TaS$_2$} heterostrcutrues}
Exfoliated \TTaS\ flakes on 90 nm SiO$_2$/Si were soaked in a 10 mM solution of iron(0) pentacarbonyl ($>$99.99$\%$, Sigma-Aldrich) in toluene. The reaction was carried out at 50~$^\circ$C for approximately 24 hours in an Ar-filled glovebox. Subsequently, the samples were rinsed three times with fresh solvent and dried under Ar gas. Following the chemical treatment, the chips were annealed at 250~$^\circ$C for 45 minutes to 1.5 hours. Typically, Raman spectra were collected after 45 minutes of annealing, and the chips were further annealed if Fe ordering was not observed.

\subsection{Preparing samples for transmission electron microscopy (TEM)} Samples for plan-view imaging (along the $c$-axis) were prepared in an Ar-filled glovebox using a dry-transfer assembly\supercite{husremovic2022hard,husremovic2023encoding}. Exfoliated flakes were picked up and placed onto 200 nm silicon nitride holey TEM grids (Norcada) that had been treated with O$_2$ plasma for 5 minutes immediately before transfer.

For cross-sectional imaging along the crystallographic $ab$-plane, lamellae were fabricated using a focused ion beam (FIB) lift-out procedure on a Thermo Fisher Scientific Helios G4 system\supercite{husremovic2022hard,husremovic2023encoding}. Prior to atomic-resolution STEM measurements, the TEM lamellae were annealed at 100–120~$^{\circ}$C in high vacuum (10$^{-7}$ Torr) for 12 hours to reduce carbon contamination.

\subsection{Transmission electron microscopy (TEM)} Atomic-resolution HAADF–STEM imaging of cross-sectional specimens was carried out on a Thermo Fisher Spectra 300 X-CFEG operated at 120 kV. A probe-forming convergence semi-angle of 24 mrad was used, with inner and outer HAADF detector collection angles of 68 and 200 mrad, respectively. Plan-view HAADF–STEM images were acquired on the TEAM I microscope (a modified FEI Titan 80–300) at 80 kV, using a beam current of 70 pA and a convergence semi-angle of 17 mrad. For cross-sectional datasets, series of fast-acquisition frames ($>$1 fps) were recorded and subsequently rigidly registered and summed\supercite{savitzky2018image} to enhance the signal-to-noise ratio while minimizing the impact of sample drift and scan noise.

Atomic-resolution HAADF-STEM images were analyzed using Fourier transform analysis to map the distribution of Fe and CDW superstructure peaks. In our analysis, we first computed the fast Fourier transform (FFT) of a real-space micrograph. Next, a single superstructure peak was selected from the FFT using a circular mask with a diameter of 0.058~\AA$^{-1}$. Subsequently, an inverse Fourier transform (iFFT) image was computed using the selected Bragg peak. The resulting iFFT is a real-space image that contains information on the spatial distribution of the superstructure fringes and their local amplitude. To suppress noise-dominated regions, the iFFT modulation amplitude was normalized by its maximum value and weighted using an intensity window corresponding to 10--20$\%$ of the maximum amplitude. The spatial distribution of superstructure fringes was then visualized after weighting by this amplitude factor.

Selected area electron diffraction (SAED) and STEM-EDS measurements were performed on an FEI TitanX operated at 60–80 kV. SAED patterns were acquired using a 40 $\mu$m diameter aperture, corresponding to a selected area of $\sim$720 nm on the sample. For STEM-EDS, data were collected with a probe convergence angle of 10 mrad and a typical beam current spanning between 150 pA and 1 nA. Total acquisition times ranged from 5 to 30 minutes, depending on the obtained signal-to-noise ratio. Data were analyzed using Bruker Esprit 1.9, and series fit deconvolution was applied to resolve overlapping peaks.

Cryogenic core-loss EELS data were acquired on an FEI Titan Themis Cryo S/TEM at an accelerating voltage of 120 kV using a Gatan 636 side-entry liquid-nitrogen holder. The microscope was equipped with a Gatan 965 GIF Quantum ER and a Gatan K2 Summit direct electron detector operated in electron-counting mode. EEL spectrum images spanning the substrate, sample, and protective surface layers were recorded in multiple regions of each specimen. Regions corresponding to the sample were then selected in Nion Swift, and the summed spectra from these regions were used for analysis\supercite{meyer2019nion}.

\subsection{Electron transport measurements}
All measurements were carried out in a Quantum Design Physical Property Measurement System (PPMS). An alternating excitation current of 1–5 $\mu$A at 17.777 Hz was applied between the source and drain contacts, and the four-probe voltage $V_{xx}$ was recorded with a Stanford Research SR830 lock-in amplifier. Magnetoresistance and Hall measurements were symmetrized using standard methods\supercite{husremovic2022hard,husremovic2025tailored}.

\subsection{Cryogenic Raman spectroscopy}
For unintercalated heterostructures, temperature-dependent Raman spectroscopy was performed using a Horiba Multiline LabRam Evolution instrument in an optical cryostat (Cryo Industries of America). A 633 nm continuous-wave laser and an ultra-low-frequency filter were used for data acquisition. Spectra were typically collected with 50–80 $\mu$W laser power and an 1800 gr/mm grating, using 10-second integration times and 2–4 accumulations. The laser was focused to a spot size of approximately 1 $\mu$m on the sample.

Cryogenic Raman measurements of sample D3 were carried out with a 532 nm diode-pumped solid-state laser (Coherent\textsuperscript{TM} Sapphire SF NX) as the excitation source. A Volume Bragg grating (VBG) based BragGrate\textsuperscript{TM} Bandpass filter is used in reflecting geometry to clean up the laser line. The laser is focused ($<$1 $\mu$m dia) onto the sample loaded on the Agile Temperature Sample mount (ATSM) in Montana Cryostation s200 with a 100x objective (NA=0.9). The Raman signal is collected in backscattering geometry and dispersed with Andor\textsuperscript{TM} Kymera 328i spectrometer using 1800l/mm grating that enables a resolution of $\sim$0.72 cm\textsuperscript{-1} at 532 nm on the detector (TE-cooled Andor\textsuperscript{TM} Newton DU970P EMCCD). To attenuate the Rayleigh scattered laser line, three VBG-based BragGrate\textsuperscript{TM} notch filters each with an OD $>$ 3 and a spectral bandwidth $<$ 5 cm\textsuperscript{-1} are aligned in series in the output beam path before the spectrometer. The excitation power on the sample was kept below 100$\mu$W to avoid any heating effects.

We note that while reference samples and D3 were measured using different lasers, there are no expected Raman shifts in the discussed modes as a consequence of the laser difference\supercite{lacinska2022raman}.

\subsection{Density Functional Theory (DFT) calculations}
We performed Density Functional Theory (DFT) calculations within the Vienna Ab Initio Simulation Package (VASP). We used projector augmented wave (PAW) pseudopotentials containing valences of $3d^64s^2$, $5p^65d^36s^2$ and $3s^23p^4$ for Fe, Ta, and S respectively. All DFT calculations were performed at the GGA+U level using the PBEsol functional and a Liechtenstein-type on-site Hubbard correction on the iron d-orbitals with U=4.0 and J=0.7. All calculations were performed using 14x14x2 and 7x7x2 $\Gamma$-centered k-point grid for a Ta$_{2}$S$_{4}$ and Ta$_{8}$S$_{16}$ slabs respectively, an energy cutoff of 600 meV, and vacuum height of 21 \textrm{\AA}. Lattice constants were either optimized to force tolerance of 0.1 meV/\textrm{\AA} for the slab unit cell at constant volume or taken to exactly match the experiment.  Experimental geometries for T and HT layers correspond to extracting bilayer interfaces from the referenced bulk 1\textit{T}\supercite{hagg1954x} and 6\textit{R}\supercite{jellinek1962system} structures without modification to form slab structures with a 21 \textrm{\AA} vacuum.  All calculations used first-order Methfessel-Paxton with $\sigma = 0.2$ eV with a negligible entropy contribution. SOC was excluded.

\section*{Acknowledgments}This research is supported by the U.S. Department of Energy (DOE), Office of Science, Basic Energy Sciences (BES), under Award $\#$ DE-SC0025525. Work at the Molecular Foundry, LBNL, was supported by the Office of Science, Office of Basic Energy Sciences, the U.S. Department of Energy under Contract no. DE-AC02-05CH11231.  Portions of this work were supported by the Heising Simons Faculty Fellowship and Philomathia Award. Electron microscopy was, in part, supported by the Platform for the Accelerated Realization, Analysis, and Discovery of Interface Materials (PARADIM) under NSF Cooperative Agreement no. DMR-2039380. This work made use of the Cornell Center for Materials Research (CCMR) Shared Facilities, which are supported through the NSF MRSEC Program (no. DMR-1719875). The Thermo Fisher Spectra 300 X-CFEG was acquired with support from PARADIM, an NSF MIP (DMR-2039380) and Cornell University. 0B.H.G. was supported by the University of California Presidential Postdoctoral Fellowship (PPFP) and Schmidt Science Fellows in partnership with the Rhodes Trust. S.H. acknowledges support from the Blavatnik Innovation Fellowship.

\section*{Conflict of Interest} The authors declare no conflict of interest.

\section*{Author Contributions}{S.H. and D.K.B. conceived the study. S.H. and W.Z. prepared the samples. S.H., B.G., K.C.B. and C.S. carried out TEM experiments. B.G. conducted the EELS measurements. S.H. and E.R.K. carried out the FFT analysis. S.H., M.D. and M.E. performed the Raman experiments. I.M.C. and S.G. conducted the DFT calculations. S.H. and D.K.B. wrote the manuscript with input from all coauthors.}

\section*{Keywords} 2D materials, magnetism, charge density waves, competing quantum phases.

\printbibliography

\clearpage

\end{document}